# Wetting of Flat Gradient Surfaces

Edward Bormashenko

*Ariel University, Engineering Faculty, Chemical Engineering, Biotechnology and Materials Department, P.O.B. 3, 40700, Ariel, Israel*

Corresponding author: Edward Bormashenko

E-Mail: Edward@ariel.ac.il

Phone: 972-3-074 7296863

**ABSTRACT**

Gradient, chemically modified, flat surfaces enable directed transport of droplets. Calculation of apparent contact angles inherent for gradient surfaces is challenging even for atomically flat ones. Wetting of gradient, flat solid surfaces is treated within the variational approach, under which the contact line is free to move along the substrate. Transversality conditions of the variational problem give rise to the generalized Young equation valid for gradient solid surfaces. The apparent (equilibrium) contact angle of a droplet, placed on a gradient surface depends on the radius of the contact line and the values of derivatives of interfacial tensions. The linear approximation of the problem is considered. It is demonstrated that the contact angle hysteresis is inevitable on gradient surfaces. Electrowetting of gradient surfaces is discussed.

1. **Introduction**

Contact angle hysteresis remains in the focus of interest of researchers working in the interface science [1-8]. Contact angle hysteresis (CAH) is inherent not only for solid/liquid but also for liquid/vapor interfaces [9-10]. Understanding of the phenomenon of contact angle hysteresis is crucial for a diversity of applications such as painting, printing and coating. Hysteresis, which is the difference between the maximum (advancing) and minimum (receding) contact angle, is caused by the adhesion hysteresis in the solid–liquid contact area (2D effect [11]) and by pinning of the solid–liquid-air triple line due to the surface roughness (1D effect) [4-8]. The phenomenological model of the CAH implying the free movement of the contact line along the substrate was suggested recently [1]. Usually contact angle hysteresis retains the motion of droplets; however, it was recently demonstrated that it may also promote the motion of floating objects [9-10].

The famous Young equation, relating the origin of the contact angle to the interaction between particles constituting a solid substrate and liquid predicts zero contact angle hysteresis (remarkably it could not be found in the famous essay by Thomas Young [12]). It asserts that the equilibrium contact angle $\theta_Y$ is unambiguously defined by the triad of the surface tensions:

$$\cos\theta_Y = \frac{\gamma_{SA} - \gamma_{SL}}{\gamma} \qquad (1)$$

where $\gamma_{SA}, \gamma_{SL}, \gamma$ are the surface tensions at the solid/air (vapor), solid/liquid and liquid/air interfaces respectively [6-8]. Various pathways of rigorous thermodynamic grounding of the Young equation were suggested, including the principle of virtual works [13], minimization of the free energy of the drop [14] and concepts supplied by non-extensive thermodynamics [15]. However, it seems that the most accurate and general derivation of the Young equation is obtained within variational treatment of the wetting problem [8, 16-19], exploiting the transversality conditions (TC) for the appropriate variational problem with free endpoints [19-20]. TC is a necessary condition for the vanishing of the first variation of a functional in the variational problems [19-20]. The presented paper demonstrates that the same variational method is applicable for the analysis of the CAH on the gradient surfaces, studied extensively in the past decade in a view of their biomedical and micro-fluidics applications [21-26]. There exist a variety of methods enabling manufacturing of gradient surfaces, one of which is the varying of the topography of the surface [23, 24]. Our paper is devoted to wetting of atomically smooth gradient surfaces, such as those reported in Refs. 22, 27, where the gradient surfaces were prepared with the diffusion controlled chemical reactions, in particular by allowing the vapor of decyltrichlorosilane to diffuse over a silicon wafer [27]. The resulting surface displayed a gradient of hydrophobicity (with the contact angle of water changing from 97º to 25º) over a distance of 1 centimeter, enabling directed transport of water droplets [22, 27].

## 2. Results and Discussion

Consider wetting of an atomically flat, gradient substrate in the situation of the partial wetting when the spreading parameter is negative [6-8]. For the sake of simplicity consider the axially symmetrical situation depicted in **Figure 1**. When a droplet is deposited on such a gradient substrate as depicted in **Figure 1**, its free energy $G$ could be written as:

$$G(h,h') = \int_o^a \left[ 2\pi \gamma x \sqrt{1+h'^2} + 2\pi x (\gamma_{SL}(x) - \gamma_{SA}(x)) + U(x,h) + 2\pi \Gamma \right] dx \qquad (2)$$

where $a$ is the contact radius, $h(x)$ is the local height of the liquid surface above the point $x$ of the substrate; $h' = \dfrac{dh}{dx}$; $\gamma = const$ is the surface tension of liquid, $\gamma_{SL}(x)$ and $\gamma_{SA}(x)$ are the $x$-coordinate-dependent solid-liquid and solid-air interfacial tensions

(recall, that the solid surface is the "gradient" one, the distribution of interfacial tensions is axially-symmetric about the *y*-axis, and the roughness of the surface is negligible), $U(x,h)$ represents the axisymmetric external field, into which the entire system is embedded, $\Gamma$ is a line tension which for the sake of simplicity is assumed to be constant across the surface [28-30], and the integral is extended over the substrate area. The first term of the integrand presents the capillary energy of the liquid cap and the second term describes the change in the energy of the gradient substrate covered by liquid. We also suppose that the droplet does not loss its mass, thus the condition of the constant volume *V* should be considered as:

$$V = \int_0^a 2\pi x h(x) dx = \text{const}. \tag{3}$$

Eqs. 2-3 reduce the problem to minimization of the functional:

$$G(h, h') = \int_0^a \tilde{G}(h, h', x) dx, \tag{4}$$

$$\tilde{G}(h, h', x) = 2\pi \gamma x \sqrt{1 + h'^2} + 2\pi x (\gamma_{SL}(x) - \gamma_{SA}(x)) + U(x, h) + 2\pi \Gamma + 2\pi \lambda x h, \tag{5}$$

where $\lambda$ is the Lagrange multiplier to be calculated from Eq. 3. For a calculation of the droplet's shape we would have to solve the appropriate Euler-Lagrange equations [19]. However, we will not calculate the droplet's shape, since our interest is the apparent equilibrium contact angle $\theta$ corresponding to the equilibrium of the droplet. Now consider that we treat the variational problem with *free endpoints* [19-20]. Thus, the TC of the variational problem should be involved [19-20]. TC at the endpoint *a* supplies:

$$(\tilde{G} - h' \tilde{G}'_{h'})_{x=a} = 0, \tag{6}$$

where $\tilde{G}'_{h'}$ denotes the $h'$ derivative of $\tilde{G}$ [16]. Substitution of Eq. 5 into the TC, given by Eq. 7, taking into account $h(a) = 0$, $U(x = a, h = 0) = 0$, and the routine transformations akin to those performed in Refs. [8, 16-17] give rise to Eq. 7:

$$\cos\theta = \frac{\gamma_{SA}(a) - \gamma_{SL}(a)}{\gamma} - \frac{\Gamma}{\gamma a}, \tag{7}$$

which looks well-expected and trivial. Indeed, the equilibrium, apparent contact angle $\theta$ depends on the values of interfacial tensions, as taken at the contact (triple) line, namely $\gamma_{SA}(a)$ and $\gamma_{SL}(a)$, as mentioned in Ref. 31. Eq.7 may be easily understood within the traditional "force-based" interpretation of the Young equation [32]. It is easily recognized that the apparent contact angle $\theta$ is independent on the external field $U(h, x)$, under assumptions adopted for this field. However Eq. 7 is less trivial than it seems from the first glance, owing to the fact that the apparent contact angle $\theta$ now depends on the radius of the contact line $a$; thus, Eq.7 immediately predicts the CAH, inevitable for gradient surfaces even when the effects due to the line tension $\Gamma$ are negligible.

Let us establish the CAH quantitatively. For this purpose we expand interfacial tensions into Taylor-McLaurin series (we restrict ourselves by the linear approximation; recall also that the surface distribution of the interfacial tensions $\gamma_{SA}(x)$) and $\gamma_{SL}(x)$ is suggested to be axisymmetric). The aforementioned expanding yields:

$$\gamma_{SA}(x) = \gamma_{SA}^0 + a\left(\frac{\partial \gamma_{SA}(x)}{\partial x}\right)_{x=0} ; \gamma_{SA}^0 = \gamma_{SA}(x=0) \tag{8a}$$

$$\gamma_{SL}(x) = \gamma_{SL}^0 + a\left(\frac{\partial \gamma_{SL}(x)}{\partial x}\right)_{x=0} ; \gamma_{SL}^0 = \gamma_{SL}(x=0) \tag{8b}$$

Substitution of Eqs.8a-8b into Eq. 7 gives rise to Eqs. 9a-9b:

$$\cos\theta = \cos\theta_0 + \frac{a}{\gamma}\left[\left(\frac{\partial \gamma_{SA}(x)}{\partial x}\right)_{x=0} - \left(\frac{\partial \gamma_{SL}(x)}{\partial x}\right)_{x=0}\right] - \frac{\Gamma}{\gamma a}, \tag{9a}$$

$$\cos\theta_0 = \frac{\gamma_{SA}^0 - \gamma_{SL}^0}{\gamma} \tag{9b}$$

Formulae 9a-9b represent the generalized Young equation for gradient surfaces. Restrict our treatment by the linear approximation for the $x$-dependencies of interfacial tensions; namely, assume:

$$\gamma_{SA}(x) = \gamma_{SA}^0 + \alpha x; \alpha = const; [\alpha] = \frac{N}{m^2} \tag{10a}$$

$$\gamma_{SL}(x) = \gamma_{SL}^0 + \beta x; \beta = const; [\beta] = \frac{N}{m^2} \tag{10b}$$

Substituting Eqs. 10a-10b into Eq. 9a immediately yields:

$$\cos\theta = \cos\theta_0 + \frac{a}{\gamma}(\alpha - \beta) - \frac{\Gamma}{\gamma a} \tag{11}$$

Formulae 9-11 express the main result of the article. The change in the apparent contact angle for gradient surfaces is governed by the derivatives of interfacial tensions; in other words: CAH hysteresis arises from physical or chemical heterogeneities of the solid, as already been suggested by Joanny and de Gennes in Ref. [33]. It is also recognized from Eqs. 9-11, that the apparent contact angle $\theta$ depends on the radius of the contact radius $a$. The precise value and the sign of the line tension $\Gamma$ are not well-known; however the values of $\Gamma$ in the range $10^{-9} - 10^{-11}$ N look realistic [28-30]; hence the effects related to line tension can be important for nano-scaled droplets or for nano-scaled rough surfaces [6-8]. For micro- and macro-scaled droplets Eq. 11 may be simplified as follows:

$$\cos\theta - \cos\theta_0 = \frac{a}{\gamma}(\alpha - \beta) \qquad (12)$$

Eq. 12 implies the size dependence of the apparent contact angle $\theta$, i.e. the contact angle hysteresis is inevitable on the gradient surfaces, with the only possible exception, occurring when the condition: $\alpha - \beta = 0$ takes place. We also conclude that the effects of gradients of interfacial tensions on the apparent contact angle become essential when the condition supplied by Eq. 13 is true:

$$\frac{a}{\gamma}(\alpha - \beta) \cong \cos\theta_0 \qquad (13)$$

Obviously, the restriction: $\left|\cos\theta_0 + \frac{a}{\gamma}(\alpha - \beta) - \frac{\Gamma}{\gamma a}\right| < 1$, arising from Eq. 11, is imposed. The deviations of the cosine of the apparent contact angle $\theta$ from the same taken in the vicinity of the geometrical center of a droplet, calculated for various numerical values of gradients of interfacial tensions and various values of $\cos\theta_0$ are supplied in **Figures 2-4**. Consider, that the sign of the difference of $\alpha - \beta$ may be negative.

The suggested approach may be easily adopted for prediction of apparent contact angles, under electrowetting [34, 35] of gradient surfaces (it was shown in Ref. 35, that electrowetting of gradient surfaces enables creation of variable focal lens controlled by an external voltage $\tilde{U}$ ). In this case the apparent contact angle $\theta$ will be supplied by Eq. 14:

$$\cos\theta - \cos\theta_0 = \frac{a}{\gamma}(\alpha - \beta) + \frac{C\tilde{U}^2}{2\gamma}, \qquad (14)$$

where $C$ is the surface capacitance.

**Conclusions**

Gradient, atomically flat surfaces are promising in a view of their micro-fluidics applications [22, 27]. Silicon wafers exposed to the vapor of decyltrichlorosilane enabled contra-intuitive upward climbing of water [27]. At the same time prediction of equilibrium contact angles on gradient surfaces is far from to be trivial, even when they are atomically smooth [8]. The generalized Young equation, derived for gradient surfaces, is reported. We conclude that the apparent (equilibrium) contact angle for droplets placed on so-called gradient surfaces [21-27] is the contact-radius-dependent value. For micro- and macro-scaled droplets (when the effects due to the line tension [28-30] are neglected), the value of the contact angle hysteresis is defined by the dimensionless value of: $\frac{a}{\gamma}(\alpha - \beta)$, where $\alpha$ and $\beta$ are the values of derivatives of interfacial tensions (suggested to be constant for the sake of simplicity); $a$ is the contact radius, and $\gamma$ is the liquid/vapor surface tension. In principle, the contact angle hysteresis on gradient surfaces may be arbitrarily large. The equilibrium apparent contact angles are independent on the external fields (gravity, etc.), acting on a droplet, placed on a gradient surface. The pure macroscopic approach, ignoring the microscopic details of the interaction of the triple line with the substrate, based on the variational treatment of the problem of wetting, should be emphasized [8]. The extension of the proposed approach to electrowetting of gradient surfaces is reported. In the future investigations it is planned to study wetting of gradient rough and chemically heterogeneous surfaces, using the variational approach, under which the contact line is free to move, exploiting the transversality conditions of the appropriate variational problem [19-20].

**Acknowledgements**

The author is indebted to Professor Gene Whyman for his longstanding fruitful cooperation in the study of wetting phenomena. I am especially indebted to Mrs. Yelena Bormashenko for her inestimable help in preparing this manuscript.

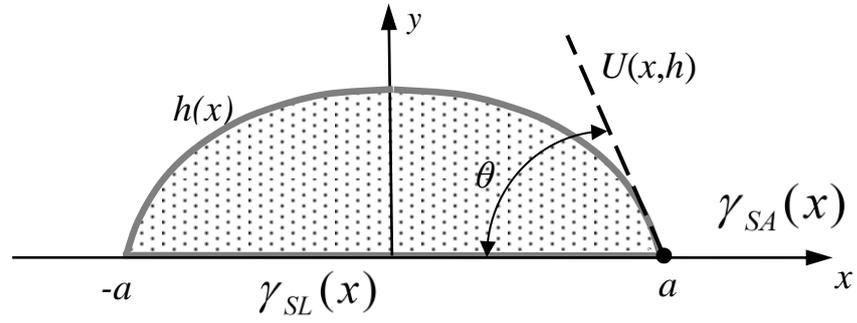

**Figure 1**. A cross-section of the spherically-symmetrical droplet deposited on the gradient substrate characterized by $\gamma_{SL}(x)$ and $\gamma_{SA}(x)$, exposed to an external axisymmetric field $U(x,h)$.

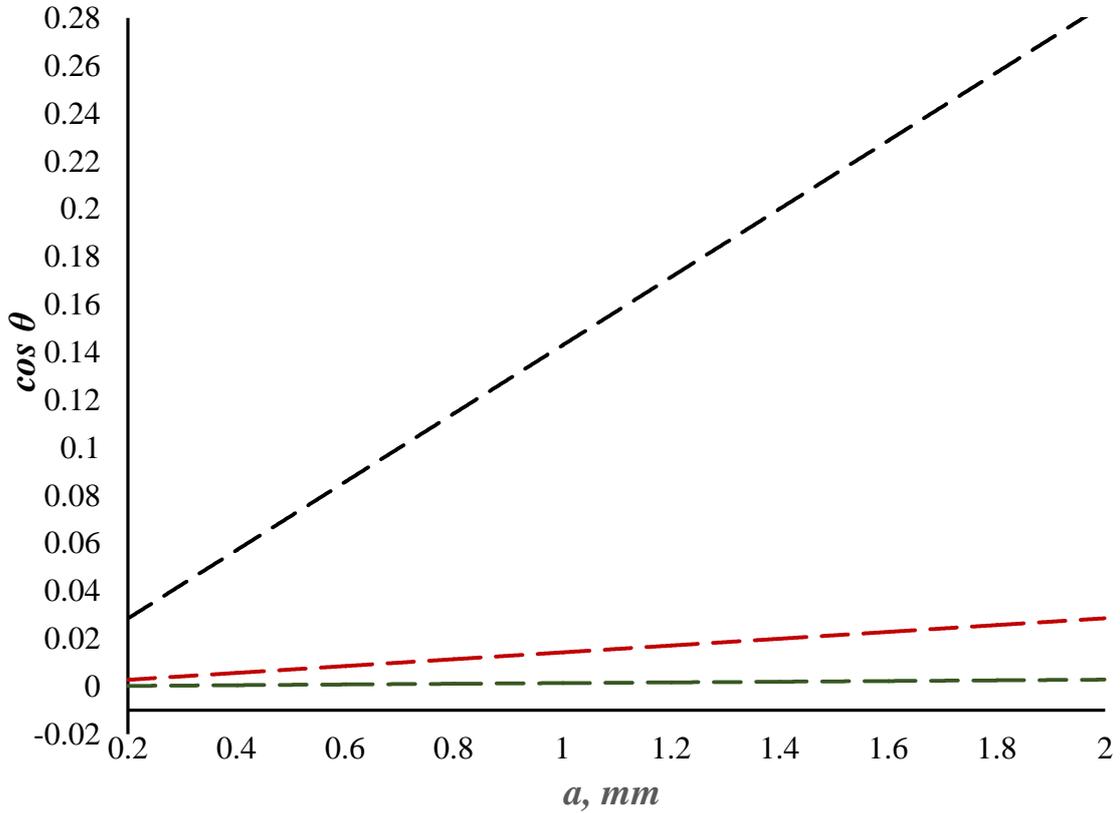

**Figure 2**. The dependence of the cosine of the apparent contact angle $\theta$ on the radius of the contact angle $a$, is plotted for water droplets ($\gamma = 70.0 \frac{mJ}{m^2}$) placed on the surface with $\theta_0 = 90^0$, as calculated with Eq. 12. The black dashed line corresponds to the difference of gradients of the interfacial tensions $\alpha - \beta = 0.1 \frac{N}{m^2}$. The red dashed line corresponds to the difference of gradients of the interfacial tensions $\alpha - \beta = 1 \frac{N}{m^2}$. The green dashed line corresponds to the difference of gradients of the interfacial tensions $\alpha - \beta = 10 \frac{N}{m^2}$.

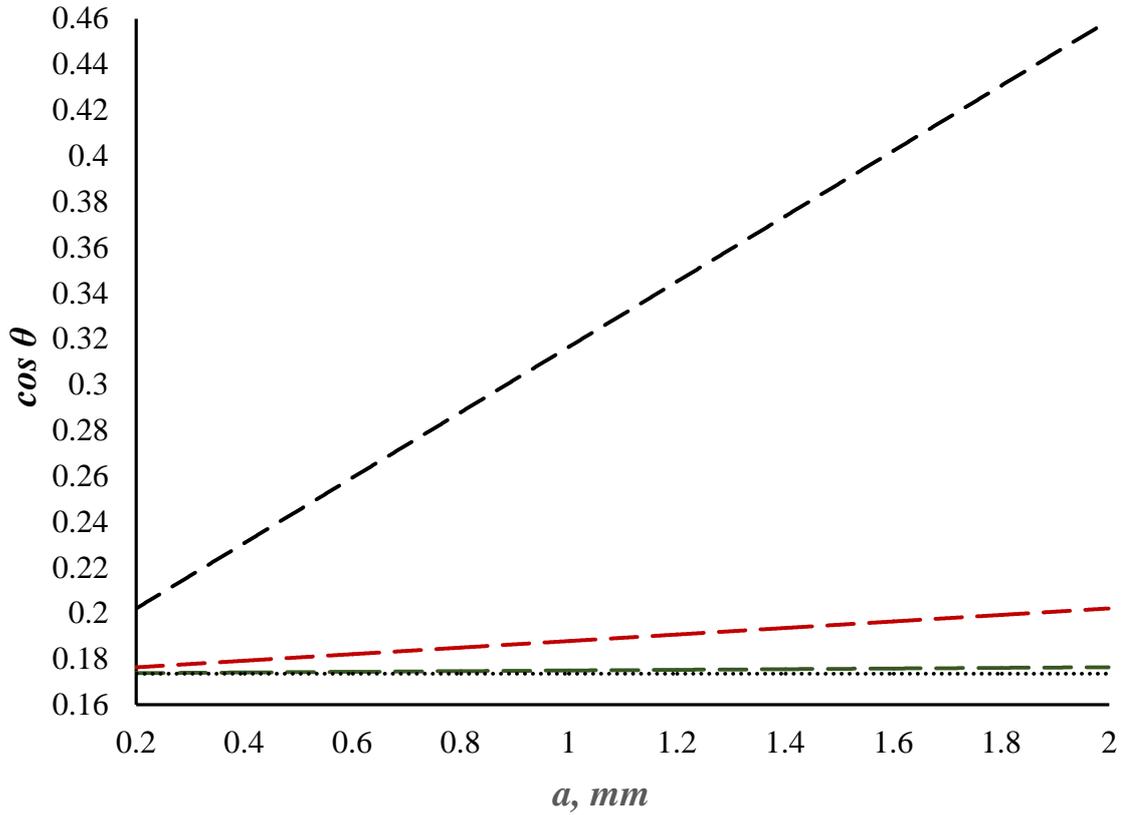

**Figure 3**. The dependence of the cosine of the apparent contact angle $\theta$ on the radius of the contact angle $a$, is plotted for water droplets ($\gamma = 70.0 \frac{mJ}{m^2}$) placed on the surface with $\theta_0 = 80^0$, as calculated with Eq. 12. The black dashed line corresponds to the difference of gradients of the interfacial tensions $\alpha - \beta = 0.1 \frac{N}{m^2}$. The red dashed line corresponds to the difference of gradients of the interfacial tensions $\alpha - \beta = 1 \frac{N}{m^2}$. The green dashed line corresponds to the difference of gradients of the interfacial tensions $\alpha - \beta = 10 \frac{N}{m^2}$. The grey dotted line corresponds to $\cos\theta = \cos\theta_0 = 0.174$.

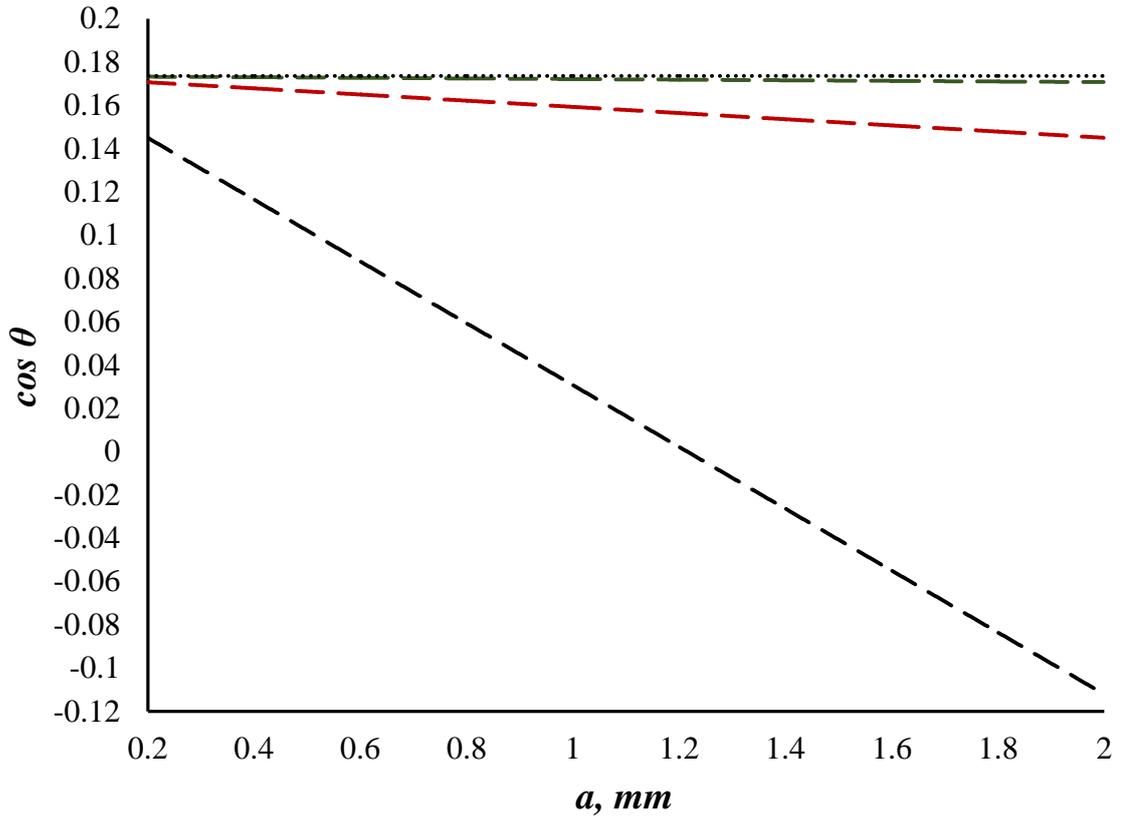

**Figure 4**. The dependence of the cosine of the apparent contact angle $\theta$ on the radius of the contact angle $a$, is plotted for water droplets ($\gamma = 70.0 \frac{mJ}{m^2}$) placed on the surface with $\theta_0 = 80^0$, as calculated with Eq. 12. The black dashed line corresponds to the difference of gradients of the interfacial tensions $\alpha - \beta = -0.1 \frac{N}{m^2}$. The red dashed line corresponds to the difference of gradients of the interfacial tensions $\alpha - \beta = -1 \frac{N}{m^2}$. The green dashed line corresponds to the difference of gradients of the interfacial tensions $\alpha - \beta = -10 \frac{N}{m^2}$. The grey dotted line corresponds to $\cos \theta = \cos \theta_0 = 0.174$.